\documentclass[letter]{aa}  

\usepackage{graphicx}
\usepackage{natbib}
\usepackage{txfonts}
\begin{document}

   \title{Resolving the stellar components of the massive multiple
     system Herschel\,36 with AMBER/VLTI }

   \author{J. Sanchez-Bermudez\inst{\ref{inst1},\ref{inst2}} \and A. Alberdi
     \inst{\ref{inst1}} \and
     R. Sch\"odel \inst{\ref{inst1}} \and
     C. A. Hummel \inst{\ref{inst2}} \and J. I. Arias \inst{\ref{inst3}} \and R.H. Barb\'a
     \inst{\ref{inst3}} \and J. Ma\'iz Apell\'aniz
     \inst{\ref{inst1},} \inst{\ref{inst5}} \and J.-U. Pott \inst{\ref{inst4}}}

      \institute{Instituto de Astrof\'isica de Andaluc\'ia (CSIC), Glorieta de la Astronom\'ia S/N, 18008 Granada, Spain. 
              \email{joel@iaa.es}\label{inst1}
         \and
        European Southern Observatory, Karl-Schwarzschild-Stra$\beta$e
        2, 85748 Garching,
        Germany. \label{inst2}
\and
        Departamento de F\'isica, Universidad de la Serena,
        Av. Cisternas 1200 Norte, 204000 La Serena, Chile \label{inst3}
\and
        Max – Planck – Institut f\"ur Astronomie, K\"onigstuhl 17,
        69117 Heidelberg, Germany \label{inst4} 
\and
Departamento de Astrof\'isica, Centro de Astrobiolog\'ia (INTA-CSIC),
campus ESA, apartado postal 78, 28\,691 Villanueva de la Ca\~nada,
Madrid, Spain \label{inst5}
}

\titlerunning{VLTI observations of Herschel\,36}
   \date{}
 
  \abstract
  {Massive stars are extremely important for the evolution of
    the galaxies; there are large gaps in our understanding of
    their properties and formation, however, mainly because they evolve rapidly,
    are rare, and distant. Recent findings suggest that most O-stars
    belong to multiple systems. It may well be that almost all massive
    stars are born as triples or higher multiples, but their large
    distances require very high angular resolution to directly
    detect the companions at milliarcsecond scales.}
  {Herschel\,36 is a young massive system located at 1.3 kpc. It has a
    combined smallest predicted mass of 45 M$_{\odot}$. Multi-epoch
    spectroscopic data suggest the existence of at least three
    gravitationally bound components.  Two of them, system $Ab$, are
    tightly bound in a spectroscopic binary, and the third one, component
    $Aa$, orbits in a wider orbit. Our aim was to image and obtain astrometric and photometric
measurements of components $Aa$ and $Ab$ using, for the first time, long-baseline optical interferometry to further constrain its nature.}
   { We observed Herschel\,36 with the near-infrared instrument AMBER
     attached to the ESO VLT Interferometer, which provides an angular
     resolution of $\sim$2 mas. We used the code BSMEM to perform the interferometric image
     reconstruction. We fitted the
     interferometric observables using proprietary IDL routines and
     the code LitPro.}
   {We imaged the $Aa$+$Ab$ components of Herschel\,36 in $H$ and $K$
     filters. Component $Ab$ is located at a projected distance of 1.81 mas, at a
     position angle of $\sim$222$^{\circ}$ east of north, the flux ratio between components $Aa$ and $Ab$ is close to one. These
     findings agree with previous predictions about the properties of
     Herschel 36. The small measured angular separation indicates that
     system $Ab$ and $Ab$ may be approaching the periastron of their orbits. These results, only achievable with long-baseline near-infrared interferometry, constitute the first step toward a thorough understanding of this massive triple system.}
   {}


   \maketitle
%

\section{Introduction \label{introduction}}


The evolution of galaxies cannot be fully understood
without studying the formation and evolution of massive
stars. They are important actors in the  enrichment and mixing of the interstellar medium. Therefore, they can initiate, accelerate, or impede the star and
planet formation. However, despite of their importance, our knowledge
about their birth and evolution is still not conclusive. This is,
mainly, because they spend a significant part of their main-sequence
lifetime (20\%) still embedded in high-extinction dust clouds, they
evolve rapidly ($\sim$ Ma), are rare, and typically located at large
distances ($\ge$ 1 kpc) \citep{Zinnecker_2007}.


Recent studies on the multiplicity of massive
stars suggest that at least 70\% of them belong to binaries or high-degree
multiple systems \citep{Mason_2009, Maiz-Apellaniz_2010, Sana_2011, Maiz_Apellaniz_2014}. Moreover, \citet{Sana_2012sci} suggested that the
interaction between components in massive binaries dominates the
evolution of massive stars. Therefore, ignoring the role of
multiplicity on the formation and evolution of massive stars introduces
strong bias in star formation theories. Despite the existing body of research, our knowledge of the multiplicity of massive stars is still incomplete, in particular (a) at the upper-end of the
initial mass function (IMF) and (b) at the early evolutionary stages
of high-mass stars (e.g., zero-age main-sequence -ZAMS-
objects). This is because observations of these systems are difficult because
of their rareness and distance \citep[e.g.,
HD\,150136;][]{Sanchez-Bermudez_2013}. Hence, studying each individual
system is necessary
to obtain reliable statistics of their properties. High-angular
resolution techniques are therefore required to characterize them fully.

\textbf{Herschel\,36} is a young massive system located at 1.3 kpc
\citep{Arias_2006} in the Hourglass high-mass star-forming region in
the central part of the M8 nebula. This object is responsible for most
of the gas ionization in the region. \citet{Arias_2010} concluded that
Herschel\,36 is composed of at least three main components: two of
them are the stars\footnote{The adopted nomenclature is consistent with the denominations of the stellar components of
Herschel\,36 according to the Washington Double Star Catalog
(WDS). \citet{Arias_2010} used $A$, $B1$, and $B2$
for components $Aa$, $Ab1$, and $Ab2$, respectively.} $Ab1$ (O9 V) and $Ab2$ (B0.5 V), which form a close
binary (system $Ab$) with a period of 1.54 days; the third one is component $Aa$ (O7.5 V), which is the most luminous star of the system
and dominates its ionizing flux. These authors also reported
strong variations in the line profile, shape, and radial velocity of
the known components of Herschel 36, which indicate that component $Aa$ and
system $Ab$ are gravitationally tied up in a wider orbit with a period of
approximately 500 days and a projected semi-major axis of 3.5 mas between
components $Aa$ and $Ab$. The smallest spectroscopic predicted mass of the system $Aa$+$Ab$ is of 19.2+26.0=45.2
M$_{\odot}$. The combined luminosity of the three components of Herschel\,36  agrees
with the expected theoretical luminosity of three ZAMS stars
\citep{Arias_2010}. This evolutionary
stage is very difficult to observe since it only lasts shorter than 1 Ma
in stars of such high mass. Moreover, in addition to the $Aa+Ab$
components, there is a fourth
star located at 0.25'' \citep[Herschel\,36 SE;][]{Goto_2006} and
other stars several arcseconds away that may also
be gravitationally bound to Herschel\,36.
Because its
location, multiplicity, and role in the nebula, this system
has therefore been compared to $\theta ^1$ Ori C \citep{Kraus_2009ori} in the
Orion Trapezium. The complete characterization of Herschel\,36
may provide new links between the observed properties and
the evolutionary models of massive stars.

Here, we report new long-baseline interferometric measurements of
Herschel\,36 with the instrument AMBER at the ESO Very Large Telescope
Interferometer (VLTI). Our
objective was resolve the two main components of  Herschel\,36, $Aa$ and system $Ab$, with the aims (i) to provide accurate astrometric measurements of their separation, (ii)
to measure their brightness ratio, and (iii) to give a first-order estimate of their orbit.

\section{Observations and data reduction \label{reduction}}

A single 1 hr observation of Herschel\,36 was obtained with AMBER \citep{AMBER_Petrov_2007}
in its low-resolution mode (LR-HK), using the VLT unit telescopes
(UTs) 1, 2, and 4 on April 17, 2014 (JD2456764.8)\footnote{Based on observations collected at the
  European Organization for Astronomical Research in the Southern
  Hemisphere, Chile, within observing programme
  091.D-0611(A). Within the same observational programme, the
    ESO archive contains additional observations of Herschel\,36 taken on April
  2013. However, these data are not
  usable for scientific purposes because of technical problems with the
  fringe tracker FINITO.}. The triplet used
for our observations has a longest baseline length of 130 m and a
shortest baseline length of 57 m. The synthesized
  beam ($\theta$=$\lambda_{min}$/2B$_{max}$) for the total JHK bandpass is 3.65 x 0.93 mas with a position
angle of 123$^{\circ}$ east of north. The instrumental setup allowed us to obtain
measurements in the $J$, $H$, and $K$ bands with a spectral resolution of
R = $\lambda$/$\Delta \lambda \sim$ 35. A standard calibrator –
science target – calibrator observing sequence was used. The chosen
calibrator, HD 165920 \citep[selected with SearchCal;][]{SearchCal1, SearchCal2}, has a diameter of 0.289 mas and is separated by 1.95$^{\circ}$ from
Herschel\,36. It is a K1 IV-V star with magnitudes of $J$=6.55,
$H$=6.13, and $K$ = 6.03, similar to the corresponding magnitudes of
the target. The airmass reported for the calibrator and target was $\sim$1.04. Figure 1 shows the u-v coverage of our observations.

For the AMBER data reduction we used the software \textit{amdlib v3}\footnote{http://www.jmmc.fr/data\_processing\_amber.htm} \citep{AMBER_Tatulli_2007, AMBER_Chelli_2009}. To eliminate
frames that were deteriorated by variable atmospheric conditions
(seeing varied between 0.6 to 1.0 at the time of our observations)
and/or technical problems (e.g., shifts in the path delay, piston), we performed
a three-step frame selection based on the following criteria: (i) first, we
selected the frames with a baseline flux higher than ten times the
noise, (ii) second, from this selection, we chose the frames with
an estimated piston smaller than 15 $\mu$m, and, finally, (iii) we
selected the 50\% of the frames with the highest signal-to-noise ratio
(S/N). The two sets of calibrator observations exhibit a similar V$^2$
response within 5\% accuracy on the UT2-UT4 and UT1-UT4
baselines. However, the squared-visibilities (V$^2$) response of the
UT1-UT2 baseline had a
larger discrepancy. Additionally, the first set of calibrator observations
presented low-photon counts. Therefore, we
decided to only use the second set of calibrator observations to
normalize our science target visibilities. The calibrated
V$^2$ and closure phases (CPs) are displayed in Fig. 2 for $H$ and $K$
bands.

\begin{figure}[htp]
\centering
\includegraphics[width=6.9cm]{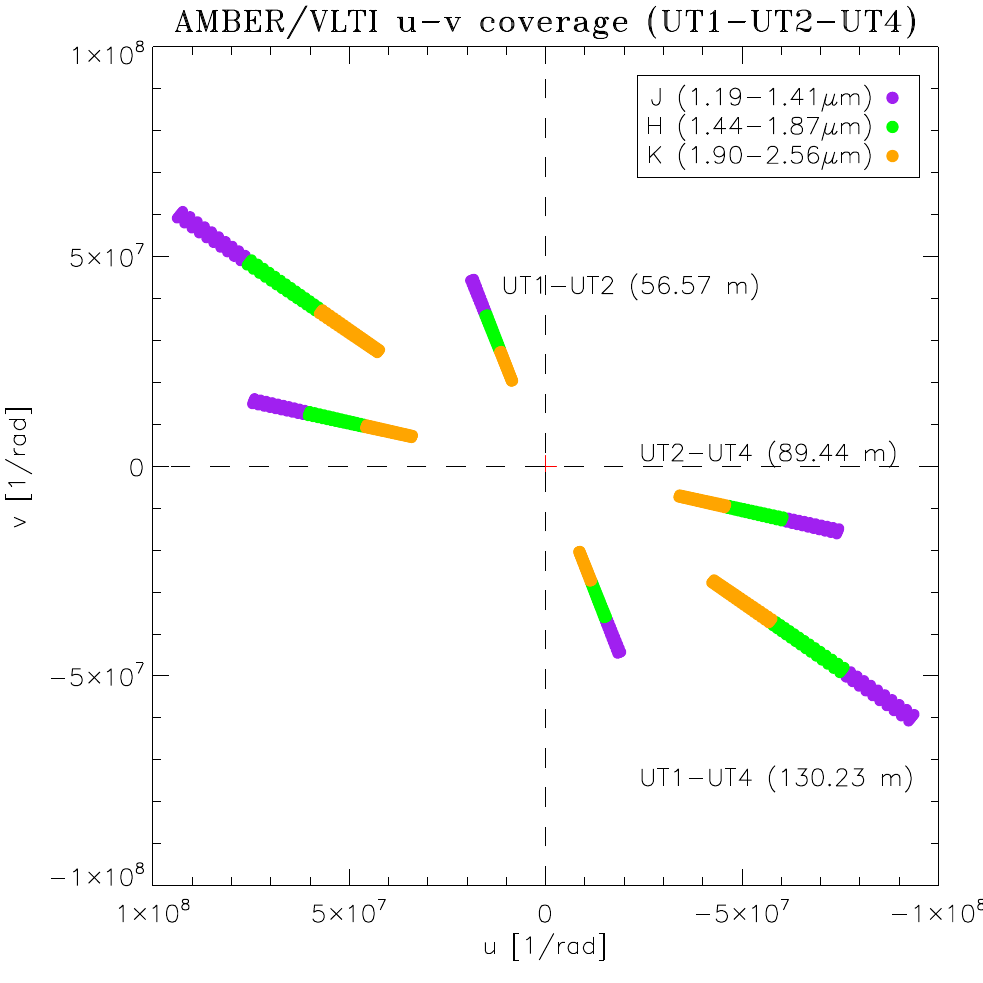} 
\caption{ u-v coverage of our AMBER/VLTI observations of
   Herschel\,36. The different bandpasses are displayed in different colors. The length of each baseline is also shown. }
 \label{fig:uv_coverage} 
\end{figure}

\section{Analysis and results \label{sec:analysis}}

\begin{figure*}[htp]
\centering
\begin{minipage}{.5\textwidth}
\centering
\includegraphics[width=6.8 cm ]{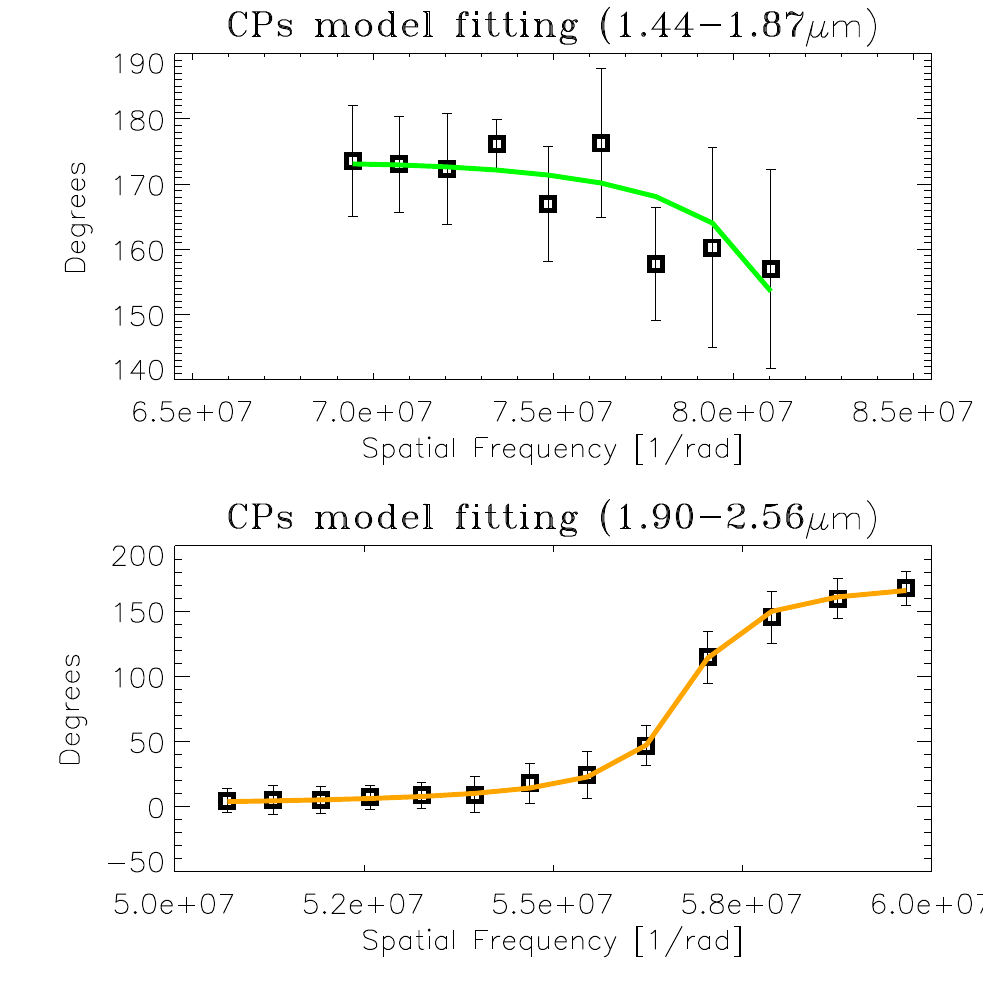}
\end{minipage}\hfill
\begin{minipage}{.5\textwidth}
\centering
\includegraphics[width=7.0 cm]{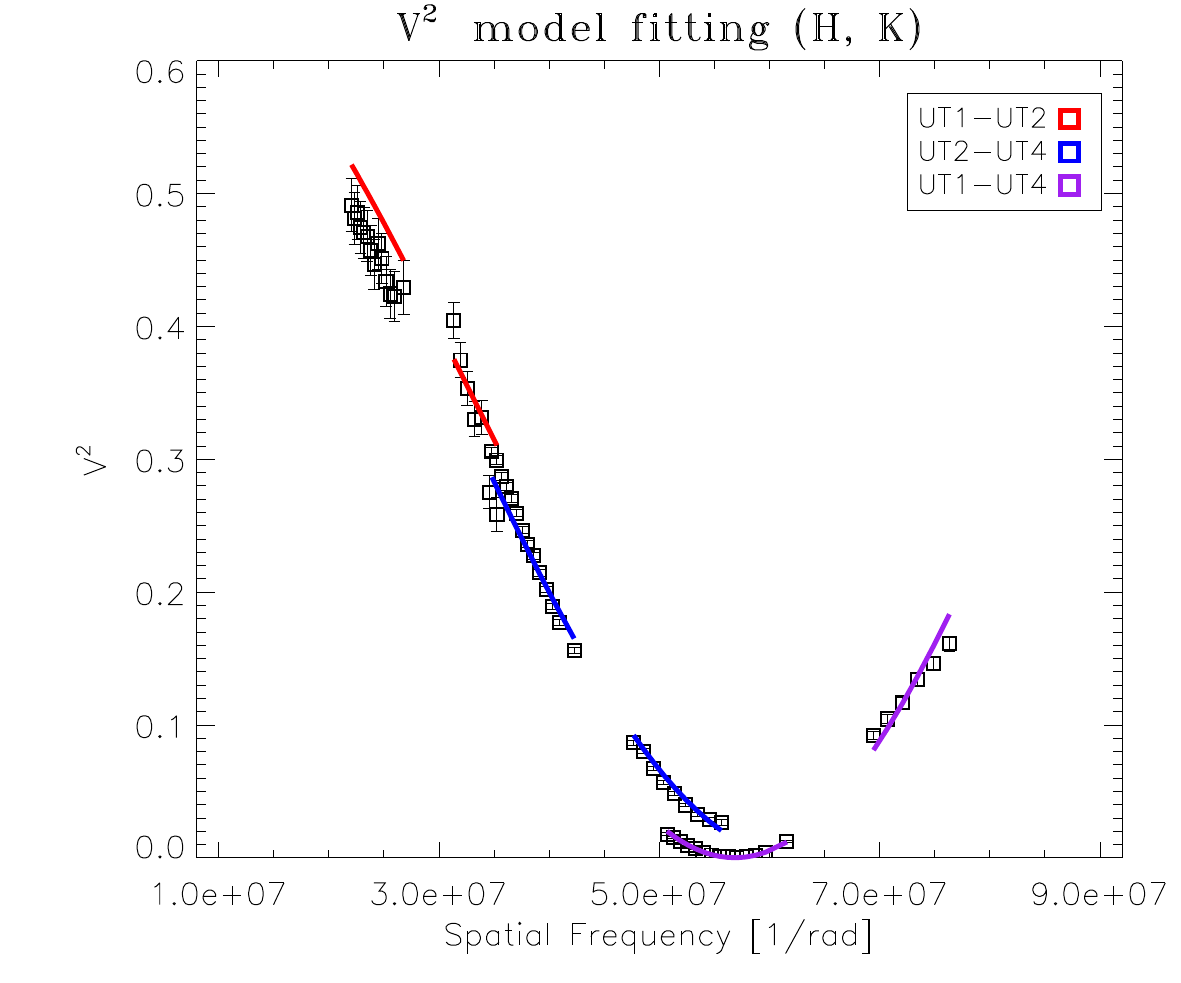}
\end{minipage}\hfill
\caption{\textit{Left:} Measured CPs of Herschel\,36. The best-fit models are represented in
  green and orange lines for the $H$ and $K$ band. \textit{Right:} Herschel\,36 V$^2$ model fitting. The
panel displays the V$^2$ for the $H$ and $K$ filters in black and the best-fit
solution at different colors for each one of the
baselines. }
 \label{fig:v2_cp} 
\end{figure*}

We used the orbital elements calculated in \citet{Arias_2010} from the
spectroscopic data of Herschel\,36 to estimate the apastron of component
$Aa$ and system $Ab$. According to the ellipse equation, the distance to
the apastron is $p=a(1+e)$, where $a$ is the semi-major axis and $e$ the
eccentricity of the system. Using the
semi-major axis, inclination, and eccentricity from Table 3 of \citet{Arias_2010}, we
estimated $p_{Aa+Ab}$ for the system $Aa$+$Ab$:

\begin{equation}
\label{eq:apoastro}
p_{Aa+Ab}=(a_{Aa}+a_{Ab})sin(i_{Aa})(1+e_{Aa+Ab})=8.12 x10^8\,km.\,
\end{equation}


For the
close binary $Ab$, we used the values from Table\,2 of
\citet{Arias_2010} to compute $p_{Ab}$:

\begin{equation}
p_{Ab}=(a_{Ab1}+a_{Ab2})sin(i_{Aa})(1+e_{Ab1+Ab2})=1.12 x10^7\,km.\,
\end{equation}


The values of $p_{Aa+Ab}$ and $p_{Ab}$ at the distance of
Herschel\,36 
thus correspond to 4 mas and  0.06 mas. In contrast
to $p_{Aa+Ab}$, the apastron of system $Ab$ indicates that it is below the VLTI angular resolution
limit at the observed frequencies. Therefore,  we fitted the calibrated V$^2$ and CPs
with a model of a binary, composed of two unresolved
sources (system $Ab$ plus component $Aa$). We used the software LITpro
to obtain the best-fit model parameters \citep{LITpro}. We emphasize that both V$^2$ and CPs exhibit a clear
cosine signature, which is typical of a binary source. 

\begin{figure}[htp]
\centering
\includegraphics[width=6.5cm]{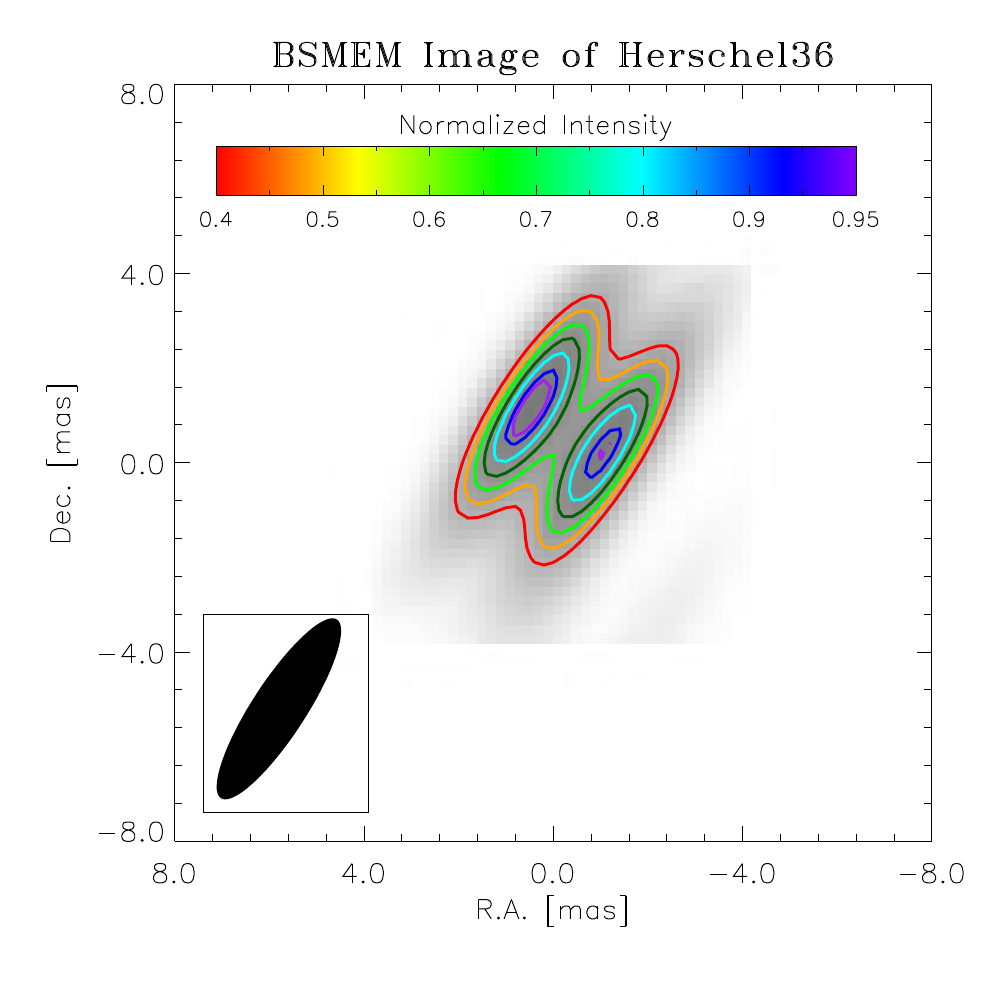} 
\caption{BSMEM image of Herschel\,36. Contours represent 40, 50, 60,
  70, 80, 90, and 95\% of the peak's intensity. The interferometric beam ($\theta$=$\lambda_{min}$/2B$_{max}$) for the data in H and K bands is of 4.50 x
1.14 mas with a
position angle of 123$^{\circ}$ east of north. The northern component
is the star $Aa$.}
 \label{fig:BSMEM_image} 
\end{figure}

 The quality of the
 $J-$band data was too low to be used in our model fitting, these data
 even introduced systematic residuals in the model
 fitting. Hence, we decided to exclude them from our
 analysis. This is expected because calibrating short
 wavelengths, such as the $J$ band, is a difficult task in optical
 interferometry. It usually requires the highest performance of the
 instrument and good atmospheric conditions. Unfortunately, there were
 seeing variations, and unstable fringe lock, and low S/N of the fringe-tracking during our observations. 

To circumvent these difficulties, we restricted the model fit to the data of the $H$ and $K$
filters. Moreover, since CPs are free of telescope-dependent phase
errors induced by atmosphere and telescope vibration, they are
usually more resistant to these systematic errors than the
V$^2$. Our analysis, showed that the V$^2$ at zero baseline of the $H$ and $K$ bands do
not reach unity. This means that an additional constant term had
to be added to our binary model to achieve a good fit to the data. This
over-resolved flux can
be explained by an additional stellar component or extended
emission (e.g., scattering halo) in the field of view ($\sim$ 60mas)
at a
larger angular distance than the one sampled by the shortest
baseline ($\sim$10 mas). To investigate these possible
scenarios in more detail, additional interferometric observations with shortest
baselines are required. The geometrical model of the $V^2$ and
CPs was thus fitted to the $H$ and $K$ filters together and to each band
independently. The best-fit model parameters are reported in
Table\,\ref{tab:CPs}. The model in Fourier space, and the
data are displayed in Fig.\,\ref{fig:v2_cp}. Any systematic
differences between
model and data are
within 2-$\sigma$ of the data uncertainties. 

Our model shows an average angular separation of 1.81$\pm$ 0.03
mas between component $Aa$ and system $Ab$. Assuming that component
$Aa$ is located at the center of the frame of reference, system $Ab$ is
located at an average projected position angle of
222$^{\circ}\pm$10.5$^{\circ}$ east of north. The best-fit flux ratio corresponds to
$f_{Ab}/f_{Aa}=0.95\pm0.12$. The $1-\sigma$ uncertainties were
estimated from the standard deviations of the best-fit values of the
fitting to the individual and the combined bands.

\begin{table}[ht]
  \caption{Best-fit of a binary model (system $Aa$ + $Ab$) of Herschel\,36
    to the V$^2$ and CPs. }
\label{tab:CPs}
\centering  
\begin{tabular}{l c c c c } 
\hline \hline                        
Parameter & Combined & $H$ & $K$ & 1-$\sigma$\tablefootmark{f}  \\ 
\hline                  
$f_{over-resolved}$\tablefootmark{a}         &  $0.17$  &   $0.18$  & $0.17$ & $0.095$\\
$f_{Aa}$\tablefootmark{b}         &  $0.42$  &   $0.42$  & $0.43$ & $0.12$\\
$f_{Ab}$\tablefootmark{c}              & $0.41$  &   $0.40$  &  $0.40$ & $0.12$ \\
$d$ [mas]\tablefootmark{d}         & $1.82$  &   $1.80$  &  $1.81$ & $0.03$ \\
$\Phi$ [deg]\tablefootmark{e}     & $234.0$  &   $214.8$  &  $217.0$ & $10.5$\\
\hline
\end{tabular}
\tablefoot{
\tablefoottext{a}{Fraction of total over-resolved flux.}
\tablefoottext{b}{Fraction of total flux contained in the tertiary ($Aa$).}
\tablefoottext{c}{Fraction of total flux contained in the inner system
  ($Ab1$+$Ab2$).}
\tablefoottext{d}{Angular separation between $Aa$ and system $Ab$ in milliarcseconds.}
\tablefoottext{e}{Angle between $Aa$ and system $Ab$ projected on the sky
  measured east of north.}
\tablefoottext{f}{1-$\sigma$ errors computed from the standard
  deviation of the best-fit models.}
}
\end{table}

Image reconstruction was performed with
the package BSMEM \citep{Buscher_1994, Lawson_2004_spie}. This code uses a maximum-entropy algorithm to
recover the real brightness distribution of the sources. To improve the quality of the image and reduce the
sidelobes, we used the
CPs and V$^2$ of both the $H$ and $K$ filters at the same time for the
image reconstruction. The best produced image was created after 54
iterations. The best image has a $\chi ^2$ of 2.16 between the data and
the final model. This
image is displayed in Fig.\,\ref{fig:BSMEM_image} and is consistent
with the best-fit geometrical model applied to the V$^2$ and
CPs. Because $f_{Aa}/f_{Ab}$ is almost one, the photometric
center is between the two components.


\section{Discussion and conclusion \label{sec:Conclusion}}

We separated for the first time component $Aa$ of Herschel\,36
from the spectroscopic binary, the system $Ab$. Our results agree
excellently well with those obtained by
\citet{Arias_2010}. The recent statistical analysis of the brightness ratio distribution in massive multiples reported by \citet{Sana_2011}
showed that close binaries typically consist of components with equal masses, while larger differences in masses can be found between components orbiting at larger angular separation. These results imply that companions at
intermediate (100-1000 mas; 100-1000 AU at 1 kpc) or large spatial
scales ($\ge$ 1''; $\ge$1000 AU at 1 kpc )
might be
formed in a process independent of the processes of close spectroscopic
binaries \citep[e.g., from a different fragmented
core and/or dynamical interactions;][]{Bonnell_2004}. 

Coplanarity of the orbits in a massive multiple systems typically indicates that
the system was formed by a single collapse event. Hence, a complete
characterization of the orbital parameters in this type of systems is
mandatory for establishing useful constraints to stellar evolutionary
models. In this respect, Herschel\,36 is an important laboratory for
determining coplanarity in massive multiples. According to
\citet{Arias_2010}, the inclinations between the orbital planes of
system $Aa$ and $Ab$ are related by the following relation: $\sin i_{Aa} = 1.03
\sin i_{Ab}$. This result restricts the inclination angles
to two possibilities: $i_{Ab} \approx i_{Aa}$ or $i_{Ab} \approx
180^{\circ}-i_{Aa}$.

\begin{figure}[htp]
\centering
\includegraphics[width=6 cm]{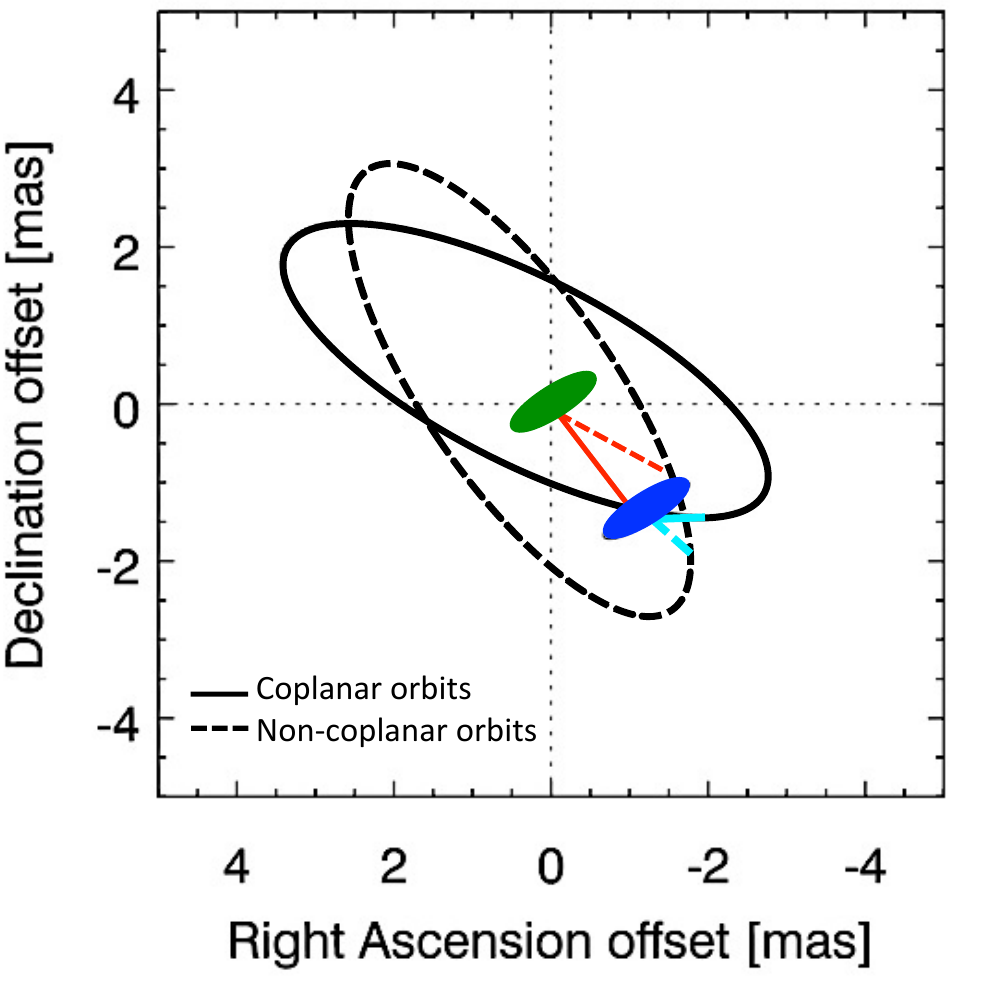} 
\caption{Orbital motion of $Ab$ (blue ellipse) around the system
    $Aa$ (green ellipse). The size of the blue and green ellipses
    corresponds to the uncertainty of the astrometric position. The
  \textit{black solid ellipse} corresponds to the orbital solution where both
  orbits are coplanar. The \textit{black dotted
    ellipse} corresponds to the non-coplanar case. The red lines
  represent the periapsis, and the light-blue lines the
distance from system $Ab$ to the projected ellipses of the orbits.}
 \label{fig:orbit} 
\end{figure}

To explore this hypothesis, we obtained a first-order
approximation of the absolute orbital parameters. To determine  a complete
solution of the outer orbit of Herschel\,36 
requires some knowledge about the orbital inclination, which can only be
derived accurately if at least three astrometric observations are
available; unfortunately, we
only have one. Therefore, some assumptions need to be made to provide
an estimate of the complete orbit. We used the estimate of the
inclination angles for the coplanar and non-coplanar cases
($i_{Aa}\sim i_{Ab} \sim$70$^{\circ}$ and $i_{Aa} \sim$ 70$^{\circ}$,
$i_{Ab}\sim $110$^{\circ}$, respectively) and the
estimate of the masses made by \citet{Arias_2010}, in addition to our estimation
of $p_{Aa+Ab}$ to perform our simulations. 



With
these assumptions, we left the angle of the ascending node ($\Omega$) as a free
parameter in our orbital simulations. The angle of the ascending node resulted in
243$^{\circ}\pm$12 for the coplanar and 214$^{\circ}\pm$50 for the
non-coplanar case. Fig.\,\ref{fig:orbit} displays the orbital
solutions. The large uncertainty in $\Omega$ can be explained as
a result of the change in the goodness of the fit due to the
measurement offset from the orbit and the projection of the ellipse onto
the line that connects the measurement and model
position. Fig.\,\ref{fig:orbit} shows that components $Aa$ and $Ab$ of
Herschel\,36 appear to be very close to their periastron passage. Using the ephemeris of \citet{Arias_2010}, we estimated the
periastrion passage for a period of 500 days of the tertiary component at the Julian date 
$T_0=2456779.5\pm8$, which is very close to the date of our AMBER
observations. The small angular separation measured with our AMBER
observations ($\sim$2 mas) compared with our apastron estimation
agrees with this prediction. 

Herschel\,36 thus belongs to the increasing number of O
stars that form hierarchical multiple systems, which can provide
important information about the high-mass star formation scenarios. To determine
whether its orbits are coplanar or not, future astrometric NIR observations with the high angular interferometric resolution of AMBER  and/or with the four-beam combiner PIONIER/VLTI in combination with a follow-up
program of the spectroscopic solution are required. 

 

\begin{acknowledgements}
We thank the referee for his/her useful comments. JSB, RS and AA acknowledge support by grants AYA2010-17631
and AYA2012-38491-CO2-02 of the Spanish Ministry of Economy and
Competitiveness cofounded with FEDER founds. RS
  acknowledges support by the Ram\'on y Cajal programme of the Spanish
  Ministry of Economy and Competitiveness. JMA acknowledges support by
  grants AYA2010-17631 and AYA2010-15081 of the Spanish Ministry of
  Economy and Competitiveness. RHB acknowledges financial support from
  FONDECYT Regular Project No. 1140076. JIA acknowledges financial
  support from FONDECYT initiation No. 11121550. JSB acknowledges support by
  the ``JAE-PreDoc'' program of the Spanish Consejo Superior de
  Investigaciones Cient\'ificas (CSIC), to the ESO studentship program.
\end{acknowledgements}
\bibliography{/Users/JSB/Documents/Papers/Paper_lib}

\end{document}